\documentclass[aps,prl,twocolumn,letterpaper,superscriptaddress,showpacs]{revtex4-1}
\usepackage[colorlinks,linkcolor=blue,anchorcolor=blue, citecolor=blue,urlcolor=blue,]{hyperref}
\usepackage{graphicx}
\usepackage{CJK}
\usepackage{verbatim}
\usepackage{physics}
\begin{document}
	
	\begin{CJK*}{UTF8}{bsmi}
		
		\title{Gauge independent description of Aharonov-Bohm effect}
		
		\author{Xiang Li (\CJKfamily{gbsn}李翔)}
		\altaffiliation{Present address: Institute for Advanced Study, Tsinghua University, Beijing 100084, China.}
		\affiliation{Tsung-Dao Lee Institute \& 
			School of Physics and Astronomy, Shanghai Jiao Tong University, Shanghai 200240, China}
		\affiliation{Zhiyuan College, Shanghai Jiao Tong University, Shanghai 200240, China}
		
		\author{Thors Hans Hansson}
		\affiliation{AlbaNova University Center, Department of Physics, Stockholm University, SE-106 91 Stockholm, Sweden}
		
		\author{Wei Ku (\CJKfamily{bsmi}顧威)}
		\altaffiliation{Corresponding email: weiku@sjtu.edu.cn}
		\affiliation{Tsung-Dao Lee Institute \& 
			School of Physics and Astronomy, Shanghai Jiao Tong University, Shanghai 200240, China}
		\affiliation{Key Laboratory of Artificial Structures and Quantum Control (Ministry of Education), Shanghai 200240, China}
		\affiliation{Shanghai Branch, Hefei National Laboratory, Shanghai 201315, China}
		
		\date{\today}
		
		\begin{abstract}
			The Aharonov-Bohm  (AB) effect is a pure quantum effect that implies a measurable phase shift in the wave function for a charged  particle that encircles a magnetic flux located in a region \textit{inaccessible} to the particle.
			Classically, such a non-local effect appears to be impossible since the Lorentz force depends on only the magnetic field at the location of the particle.
			In quantum mechanics, the Hamiltonian, and thus the Schr\"odinger equation, has a local coupling between the current due to the particle, and the electromagnetic vector potential $\mathbf{A}$, which extends to the entire space beyond the region with finite magnetic field.
			This has sometimes been interpreted as meaning that in quantum mechanics $\mathbf{A}$ is in some sense more ``fundamental" than $\mathbf {B}$ in spite of the former being gauge dependent, and thus unobservable. 
			Here we shall, with a general proof followed by a few examples,  demonstrate that the AB-effect can be fully accounted for by considering only the gauge invariant $\mathbf{B}$ field, as long as it is included as part of the quantum action of the entire isolated system. 
			The price for the gauge invariant formulation is that we must give up locality -- the AB-phase for the particle will arise from the change in the action for the $\mathbf{B}$ field in the region inaccessible to the particle.
		\end{abstract}
		
		\maketitle
	\end{CJK*}
	
	\section{ I. Introduction}
\begin{figure}
		\setlength{\belowcaptionskip}{-0.5cm}
		\vspace{-0.1cm}
		\begin{center}
			\includegraphics[width=\columnwidth]{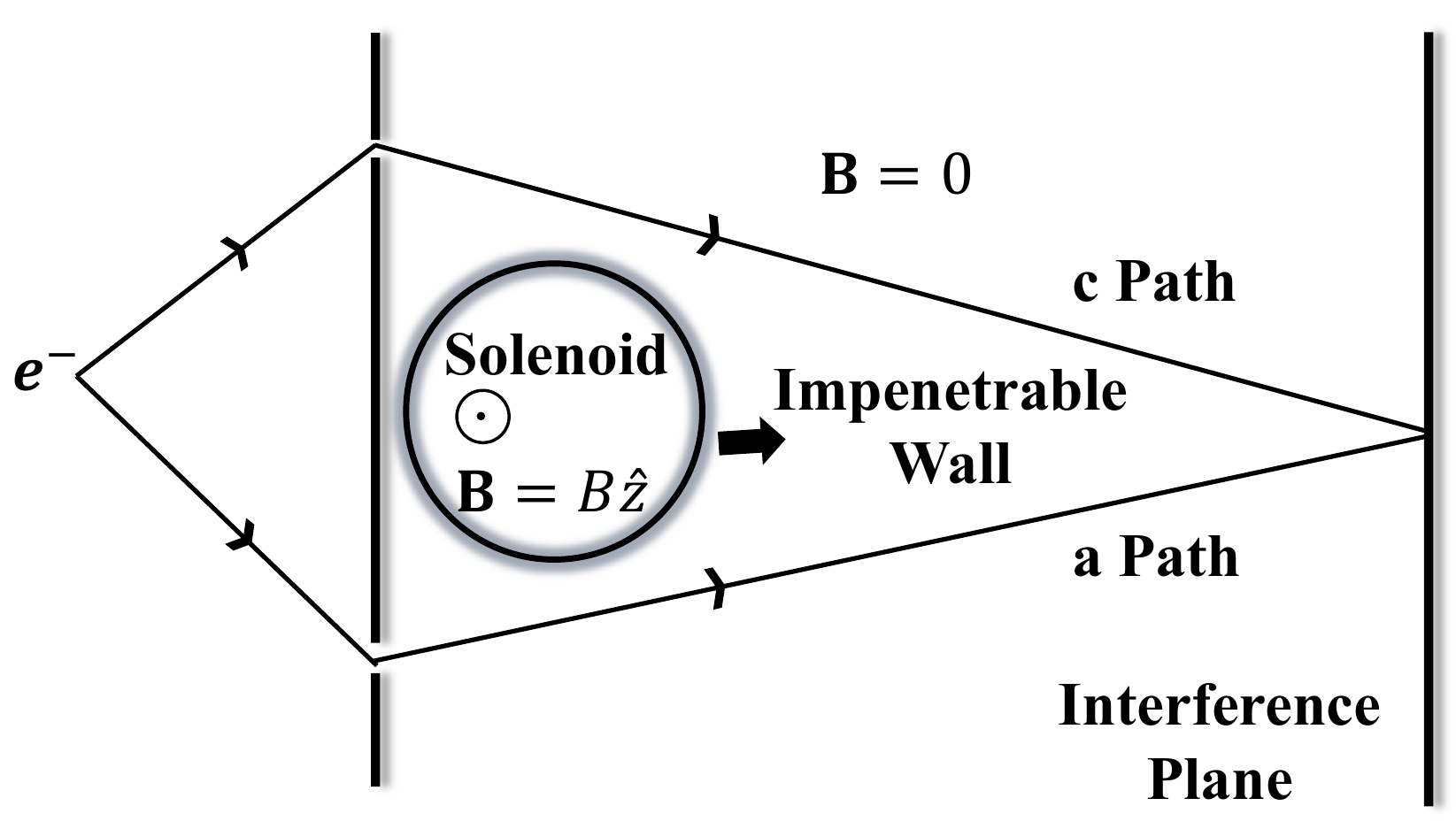}
			\caption{The Aharonov-Bohm effect illustrated by a two-slit experiment in the presence of a magnetic flux localized inside an infinitely long solenoid. The magnetic field is strictly  zero in the whole region where the charged particles moves, but the interference pattern on the screen is still influenced by the flux. Shown are two paths passing in a clockwise direction(c), and an anti-clockwise (a) direction around the solenoid. }
			\label{fig1}
		\end{center}
	\end{figure}

	Proposed in 1959\cite{PhysRev.115.485}, and experimentally confirmed  some years later~\cite{PhysRevLett.5.3} the Aharonov-Bohm (AB) effect describes a phase shift of a charged quantum  particle due to the presence of magnetic field in a region \textit{inaccessible} to the particle.
	Figure~\ref{fig1} illustrates the paradigmatic set-up with a magnetic field confined inside a tightly wound solenoid shielded by an impenetrable wall.
	Since the  particle cannot move across the wall, it  never experiences any Lorentz force, and by intuition based on classical electromagnetism, one would expect the field to have no effect at all.
	Surprisingly, using only  the canonical formulation of a quantum mechanical particle coupled to  electromagnetism, Aharonov and Bohm showed~\cite{PhysRev.115.485} that the wave function of the  particle would still acquire a phase shift depending on the 
	magnetic flux $\Phi_\mathbf{B} = \int_{\text{inside}}d\mathbf{S}\cdot\mathbf{B}$. This phase shift can be experimentally observed as a flux dependence in the interference patterns.
	
	The textbook explanation of the AB effect suggests that in quantum mechanics  one \emph{must} introduce a  vector potential in order to couple charged particles to electromagnetism, while in classical theory only the fields are necessary. 
	Outside the solenoid, even though the magnetic field is zero, a charged quantum particle still experiences a non-zero vector potential $\mathbf{A}$, whose loop integral $\oint d\mathbf{l}\cdot\mathbf{A} $ around the magnetic field gives the total magnetic flux $\Phi_\mathbf{B}$ inside the solenoid.
	The presence of a flux can thus give an observable effect even when the charged quantum particle moves only in a field-free region.
	
	The Aharonov-Bohm effect has had a profound impact on the common understanding of the role of the vector potential in quantum physics, and it is a wide spread notion that  the vector potential is more ``fundamental'' than the magnetic field.
	Such a position is clearly not very satisfactory, since $\mathbf{A} $ does not have a direct physical meaning -- it can be changed by a gauge transformation, $\mathbf{A}\rightarrow\mathbf{A}+\gradient \lambda$.
	The sharp question is: What is the minimal information needed in order to describe how a charged particle couples to an electromagnetic field?
	This was answered in a classical paper by 1965 by Wu and Yang~\cite{PhysRevD.12.3845}, in which they showed that what is needed is the set of all the gauge invariant phase factors,
	\begin{equation} \label{wilson}
	W[C] = e^{\frac{iq}{h} \oint  d\mathbf{l}\cdot\mathbf{A}} = e^{ \frac{iq}{h} \Phi_\mathbf{B}}
	\end{equation}
	where the integral is along an arbitrary closed loop $C$ and $q$ is the charge of the particle.
	If the particle moves in a simply connected region, meaning that any loop can be continuously contracted to a point, the line integral in Eq.~\eqref{wilson} can be written as a surface integral over the magnetic flux, and thus can be calculated from the magnetic field.
	In a non-simply connected region, there are loops that cannot be contracted, and  to compute the value of these Wilson loops, extra information is required.
	The AB set up in Fig.~\ref{fig1} is the simplest example.
	Note that here it is sufficient to give the value of  $W[C]$ for a single (arbitrary) loop that encircles the solenoid, since any other non-contractible loop can be reconstructed from this one again using Stokes's theorem and the knowledge of $\mathbf{B}$.
	In this simple example, it is also clear that to calculate $W[C]$ for one of the non-contractible loops, it is sufficient to know the flux through the solenoid.
	For more complicated geometries it is still true that what is needed to encode all electromagnetic effects is the field strengths together with a number of Wilson loops, or equivalently, a set of fluxes in the non-accessible regions.
	To summarize, although we only need to specify certain line integrals and do not need to specify $\mathbf{A}$, it nevertheless appears that the vector potential is needed in order to give a \emph{local} description of the interaction between the charge particle and the electromagnetic field. 
	
	Recently, the nature of the AB effect has again attracted attention, and several recent theoretical studies have appeared.
	Vaidman attempted to explain the AB phase without gauge-dependent potentials by a \emph{local force} on the source of the potential~\cite{PhysRevA.86.040101}.
	However, this force-based explanation, and the claim that the non-locality can be eliminated, was challenged by Aharonov and collaborators, who gave six examples where the AB effect is still present without there being any forces on the source, and showed non-locality is still needed~\cite{PhysRevA.92.026101}.
	Meanwhile, Kang analyzed a two-particle system consisting of a charge and a flux separated by different kinds of shields. 
	His  analysis also concluded that a local field interaction is needed for a universal description~\cite{PhysRevA.91.052116}.
	After that, Aharonov and collaborators~\cite{PhysRevA.93.042110} re-emphasized the necessity of non-locality to get a proper description of the AB effect.
	They further claimed that a gauge-dependent description is needed  to handle instantaneous interactions.
	Heated discussions continued.
	Using a gauge dependent analysis, Pearle and Rizzi~\cite{PhysRevA.95.052123, PhysRevA.95.052124} showed that the AB phase can be attributed to one of three entities: the electron, the solenoid, or  the vector potential.
	Marletto and Vedral~\cite{PhysRevLett.125.040401} pointed out a residual non-locality in Vaidman's model. 
	and proposed that it can be mediated by a local coupling between the charges and the quantized electromagnetic field.
	They claimed that the AB phase is generated \emph{locally}, being built up gradually along the path of the charge.
	Even more recently
	Saldanha~\cite{PhysRevA.104.032219, Saldanha}, proposed a description of  the AB effect through local interactions of charges and currents with a quantized electromagnetic field.
	He attributed the AB phase to a correction to the electromagnetic vacuum energy due to the local interaction.
	In summary, no consensus seems to have been reached concerning two essential aspects of the AB effect: 1) the possibility of a gauge-invariant description, and 2) the necessity of non-locality.
	
	In this paper we provide a different perspective to this discussion.
	Instead of taking just the moving particle (or particles) as the quantum system, we shall expand it to include also the electromagnetic fields.
	We will thus include the term $-\int dt\, E_F(t)$, where $E_F(t) $ is the energy in the electromagnetic field, and then show that the action due to minimally coupling the charged quantum particle to the  potentials can  alternatively be viewed  as the change in the action of the external field induced by the moving  particle.
	Using the path integral formulation of quantum mechanics, this  naturally leads to an additional phase in the path integral that depends \emph{only} on the gauge-invariant electromagnetic field.
	In two special cases, followed by a general case, we will demonstrate that this additional phase is precisely the AB phase shift that can also be obtained from  the vector potential. 
	Our calculation sheds light on the two questions posed above.
	It demonstrates that the AB effect \emph{can} be explained using only gauge invariant quantities, and also that doing so necessarily involves non-local effects.
	As we shall show, the latter appears, since the change in the field action comes only from the region inside the solenoid, and thus is arbitrarily far from the moving particle. 
	After these examples we discuss the electric version of the AB effect, using a thought experiment due to Vaidman~\cite{PhysRevA.86.040101}. 
	We end the paper by first making a comparison to the earlier works mentioned above, and then giving a summary and pointing out some open questions.
	
	
	\section{II. AB effect -- the local picture}
	
	In this section, we first give a brief account of the  conventional way of describing the AB effect in the Lagrangian version of quantum mechanics.
	Consider a charged quantum particle in the presence of electromagnetic potentials $\mathbf {A}$ and $A_0$,
	\begin{eqnarray}
	\label{Le+LeF}	L_p=L_{0} + L_{int} = \frac{1}{2}m\Dot{\mathbf{x}}^2+[q\Dot{\mathbf{x}}\cdot\mathbf{A(x)} -qA_0(\mathbf{x})]
	\end{eqnarray}
	Here $L_0$ denotes the kinetic energy, and $L_{int}$ is the \emph{local} coupling between the particle and the field via the vector potential $\mathbf{A}(\mathbf{x})$ and the scalar  potentials $A_0(\mathbf{x})$.
	One can now understand why the particle can still be influenced by the field in Fig.~\ref{fig1}:  $\mathbf{A}(\mathbf{x})$ is non-zero even in the region with zero $\mathbf{B}$ field.
	
	Using a path integral formulation, we can  easily calculate how the flux in the solenoid influences the interference pattern on the interference plane in  Figure.~\ref{fig1}. The amplitude at a point $\mathbf{x}_f $ on the screen is given by $A_{c} + A_{a}$ ($c$ indicates clockwise and $a$ indicates anti-clockwise) where
	\begin{align} \label{pathsint}
		A_{c/a} = \int_{\mathrm {c/a \ paths}} {\mathcal D}[\mathbf {x}(t)] e^{\frac{i}{\hbar}S[\mathbf {x}(t)]}
	\end{align}
	where $S = \int_0^T dt\, L_p$ is the action, and the sums are over all paths starting from $\mathbf {x}_i$ at $t=0$ to $\mathbf{x}_f$ at $t=T$, which run in the clockwise and anti-clockwise directions, respectively.
	The probability of finding the particle at point $\mathbf{x}_f$ is given by $|A_{a} + A_{c}|^2$ so the interference  is extracted from the cross term $A_c^* A_a + A_a^* A_c$. Substituting Eq.~\eqref{Le+LeF} into Eq.~\eqref{pathsint} we can immediately extract the flux dependence from,
	\begin{eqnarray}
	\begin{aligned} \label{phasefactor}
	A_a^* A_c \sim e^{\frac{i}{\hbar}(S_c-S_a)} &\sim e^{\frac{iq}{\hbar}(\int_cdt-\int_adt)\, \dot{\mathbf{x}}\cdot\mathbf{A}}\\
	&=e^{\frac{iq}{\hbar}(\int_cd\mathbf{x}-\int_ad\mathbf{x})\,\cdot\mathbf{A}}=e^{\frac{iq}{\hbar}\oint  d\mathbf{l}\cdot\mathbf{A}} 
	\end{aligned}
	\end{eqnarray}
	where we recognize the phase factor Eq.~\eqref{wilson}, and we recall that the integral is along an arbitrary loop that encloses the solenoid once in the positive direction. 
	
	We again stress that,  in this  approach, to have a local coupling to the particle, one must introduce the vector potential $\mathbf{A}(\mathbf{x})$, but there is no local description that involves only the gauge invariant field $\mathbf{B}$.  
	
	Before turning to the AB effect in the next section we use a very simple and well-known example to remind ourselves that a physical quantity, here the electrostatic energy, can be accounted for in different ways.
	For this, consider the electrostatic energy of two  charges $q_1$ and $q_2$, at positions $\mathbf{ r}_1$ and $\mathbf{ r}_2$ corresponding to the charge distribution, $\rho_i(\mathbf {x}_i)=q_i \delta^3 (\mathbf {x}_i - \mathbf {r}_i), i=1,2$ respectively:
	\begin{align} \label{heuren1}
		E_{C} &=    q_1 A_0^{(2)} (\mathbf r_1) 
		=  \int d\mathbf{ x}_1\,  \rho_1(\mathbf {x}_1) A_0^{(2)}(\mathbf{x}_1)   \\
		&=  \epsilon_0\int d\mathbf {x}_1\, \gradient\cdot\mathbf {E}_1(\mathbf {x}_1) A_0^{(2)} (\mathbf {x}_1)   \label{heuren3}\\
		\label{heuren4}
		&=  \epsilon_0\int d\mathbf {x}\, \left[ \mathbf{E}_1 \cdot \mathbf{E}_2 \right] (\mathbf {x}) \\
		\label{heuren5}
		&= \frac {\epsilon_0} {2} \int d\mathbf {x}\, \left[ \mathbf {E}^2 - \mathbf{E}_1^2 - \mathbf{E}_2^2 \right] (\mathbf {x}) \, ,
	\end{align}
	where in the first line the energy is that of the charge $q_1$ in the gauge dependent electrostatic potential due to the charge $q_2$ at position $\mathbf r_2$.
	Note that in this case the gauge dependence is just the freedom to pick a zero for the electrostatic energy. 
	Equation \eqref{heuren3} follows from Gauss's law, $\gradient\cdot \mathbf{E} = \rho/\epsilon_0$, and Eq.~\eqref{heuren4} by an integration by parts and using $\mathbf{E} = -\nabla A_0$. 
	Equation \eqref{heuren5} is the energy stored in the total electric field $\mathbf{E} = \mathbf{E}_1 + \mathbf{E}_2$, where we subtract the self energies to retain only the interaction energy. Thus the local expression Eq.~\eqref{heuren1} involving the gauge dependent quantity $A_0$ can be traded for a non-local expression in terms of the gauge invariant electric field. 
	As we shall show later, the identities Eq.~\eqref{heuren1} to Eq.~\eqref{heuren5} will be central to understand the so-called ``electric AB effect''  discussed in Refs.~\cite{PhysRev.115.485, PhysRevA.86.040101, PhysRevA.95.052123, PhysRevA.95.052124, PhysRevA.104.032219, Saldanha}.
	
	
	\section{III. AB effect -- the non-local picture} \label{III}
	
	Since the electrostatic energy is part of the Lagrangian for the electromagnetic field, it is natural to ask whether some  rewriting similar to that done above can be used to express the AB phase shift in terms of only magnetic field, and we now demonstrate that this is indeed the case.
	Some care is needed to get the signs correct, which should come as no surprise to those familiar with calculating magnetic energies.
	We also include the electric energy discussed above to get a covariant expression.
	
	As  mentioned in the Introduction, the main idea  is to include the electromagnetic field as part of the \emph{quantum} system, and calculate the flux dependence of the phase acquired when the particle encircles the solenoid. 
	The Lagrangian for the entire system will, in addition to $L_{0}$ describing the particle, include the term $-\int dt\, E_F(t)$, where $E_F$ is the energy stored in the electromagnetic field generated by both the solenoid, and the moving particle.
	Here we assume that the particle moves so slowly that we can neglect radiation effects, and that the current in the solenoid is constant, so the only time dependence comes from the field generated by the particle.
	With this approximation we get the expression,  
	\begin{eqnarray}
	\label{LF}
	\begin{aligned}
	L_{F} &= \int d\mathbf {x^\prime} \, \left(\frac{\mu_0}{2}  \mathbf {H}^2-\frac{1}{2\epsilon_0} \mathbf {D}^2\right) \\
	&=- E_F = \int d\mathbf {x^\prime} \, \left(\frac{1}{2\mu_0}  \mathbf {B}^2-\frac{\epsilon_0}{2} \mathbf {E}^2\right),
	\end{aligned}
	\end{eqnarray}
	where $\mathbf D = \epsilon_0\mathbf{E}$ is the electric displacement, and $\mathbf H = \frac{1}{\mu_0}\mathbf{B}$. 
	The energy density has two parts, the electric one, related to the longitudinal electric field, which we already encountered in Eq.~\eqref{heuren5}, and the magnetic energy density, ${\mathcal E}_{mag} = -  \frac{1}{2\mu_0} B^2$  where $\mathbf{B}$ is related to the classical current distribution by Maxwell equations.
	The origin of the non-intuitive minus sign is explained in many textbooks (for an authoritative exposition, see Ref.~\cite{landau1984electrodynamics}) and also in the Appendix.
	The basic point is that we consider the sources as given, not the fields. 
	
	We now rewrite the interaction term in the particle Lagrangian as, 
	\begin{eqnarray}
	\label{model1}
	\begin{aligned}
	L_{int}&=q\, \Dot{\mathbf{x}}\cdot\mathbf{A(x)}-q\, A_0(\mathbf{x})\\
	&=\int d^3\mathbf{x^\prime}\left[q\, \delta(\mathbf{x^\prime}-\mathbf{x})\mathbf{\dot{x}}\cdot\mathbf{ A(\mathbf{x^\prime})}-q\, \delta(\mathbf{x^\prime}-\mathbf{x})A_0(\mathbf{x^\prime})\right]\\
	&=\int d^3\mathbf{x^\prime}\left[\left(\gradient^\prime\times\delta\mathbf{H}\right)\cdot\mathbf{A}-\left(\gradient^\prime\cdot\delta\mathbf{D}\right)A_0\right]\\
	&=\int d^3\mathbf{x^\prime}\left[\frac{1}{\mu_0}\left(\gradient^\prime\times\delta\mathbf{B}\right)\cdot\mathbf{A}-\epsilon_0\left(\gradient^\prime\cdot\delta\mathbf{E}\right)A_0\right]\\
	&=\int d^3\mathbf{x^\prime}\left[\frac{1}{\mu_0}\delta\mathbf{B}\cdot(\gradient^\prime\times\mathbf{A})+\epsilon_0\delta\mathbf{E}\cdot\gradient^\prime A_0\right]\\
	&=\int d^3\mathbf{x^\prime}\left(\frac{1}{\mu_0}\delta\mathbf{B}\cdot\mathbf{B}- \epsilon_0\delta\mathbf{E}\cdot\mathbf{E} \right)(\mathbf{x},\mathbf{\dot{x}})\\
	&=\int d^3\mathbf{x^\prime}\left(\frac{1}{2\mu_0}(\delta F)_{\mu\nu}F^{\mu\nu} \right)(\mathbf{x},\mathbf{\dot{x}})\, ,
	\end{aligned}
	\end{eqnarray}
	where, $\delta\mathbf{E}(\mathbf{x^\prime};\mathbf{x})=\frac{1}{\epsilon_0}\delta\mathbf{D}(\mathbf{x^\prime};\mathbf{x})$ and $\delta\mathbf{B}(\mathbf{x^\prime};\mathbf{x},\dot{\mathbf{x}})=\mu_0\delta\mathbf{H}(\mathbf{x^\prime};\mathbf{x},\dot{\mathbf{x}})$ denote the fields at $\mathbf{x^\prime}$ 
	induced by the  charged particle at $\mathbf{x}$ moving with velocity $\dot{\mathbf{x}}$, while $\mathbf E$ and $\mathbf B$ refer to the electromagnetic fields due to external charge and current distribution.
	The last line of Eq.~\eqref{model1} is a relativistically covariant form, where $F_{\mu\nu}=\partial_\mu A_\nu-\partial_\nu A_\mu$, and where we define $A^\mu=\left(\frac{A_0}{c},\mathbf{A}\right)$ with $c=\frac{1}{\sqrt{\epsilon_0\mu_0}}$ being speed of light.
	
	When specialized to the AB case shown in Fig.~\ref{fig1}, $\mathbf B$ is the field induced by the current in the solenoid, and $\mathbf E = 0$. We get 
	\begin{eqnarray}
	\label{model2}
	\begin{aligned}
	L_{int}
	&=\int d^3\mathbf{x^\prime} \frac{1}{\mu_0}\delta\mathbf{B}\cdot\mathbf{B}  
	= \delta \int d^3\mathbf{x^\prime} \frac{1}{2\mu_0}  \mathbf B_{tot}^2  \\
	&= \delta L_{F}(\mathbf{x},\mathbf{\dot{x}}) \, ,
	\end{aligned}
	\end{eqnarray}
	where $\mathbf B_{tot} = \mathbf B + \delta \mathbf B$, and the second equality again holds in limit of vanishing velocity where the $\delta\mathbf B \cdot \delta\mathbf B$ term can be neglected.
	
	Since $\delta L_{F}$ equals $L_{int}$, their contributions to the total action, and thus to  the path integral equation.~\eqref{pathsint} are the same, which proves that the AB phase shift can be calculated using either of these expressions. In addition to this topological phase that depends on the flux in the solenoid, there is  also a dynamical phase which will, however, not have any flux dependence. 
	Thus, we have achieved our goal to provide a derivation of the AB phase using only the electromagnetic field, but at the expense of including the field as part of the quantum action as a non-local effect.
	
	
	\section{IV. The Magnetic AB Effect}
	In this section we illustrate the above result with several case studies of the magnetic AB effect in various  simple geometries.
	\subsection{An infinite  solenoid}
	Let us first consider the special case illustrated in Fig.~\ref{fig1}  which was discussed in the original paper by Aharonov and Bohm. We take a thin cylindrical solenoid of infinite length, which is very closely wound so that the magnetic field is fully confined within the solenoid.
	As shown in the previous section, the action in the presence of the magnetic field can be obtained from that in the absence of the  field, denoted by $S_{e}$, as:
	\begin{eqnarray}
	\label{Sreplace}
	S_{e} \rightarrow S_{e}+\frac{1}{\mu_0}\int dt\int d^3\mathbf{x^\prime \;B}\cdot\delta\mathbf{B}
	\end{eqnarray}
	So the action difference between paths going clockwise  and anti-clockwise is just the action of a path turning one circle.
	\begin{eqnarray}
	\label{deltaS}
	S_{c}-S_{a}=\Delta S = \frac{1}{\mu_0}\int_0^T dt\int d^3\mathbf{x^\prime\; B}\cdot\delta\mathbf{B}
	\end{eqnarray}
	
\begin{figure}
		\setlength{\belowcaptionskip}{-0.5cm}
		\vspace{-0.1cm}
		\begin{center}
			\includegraphics[width=6.5cm]{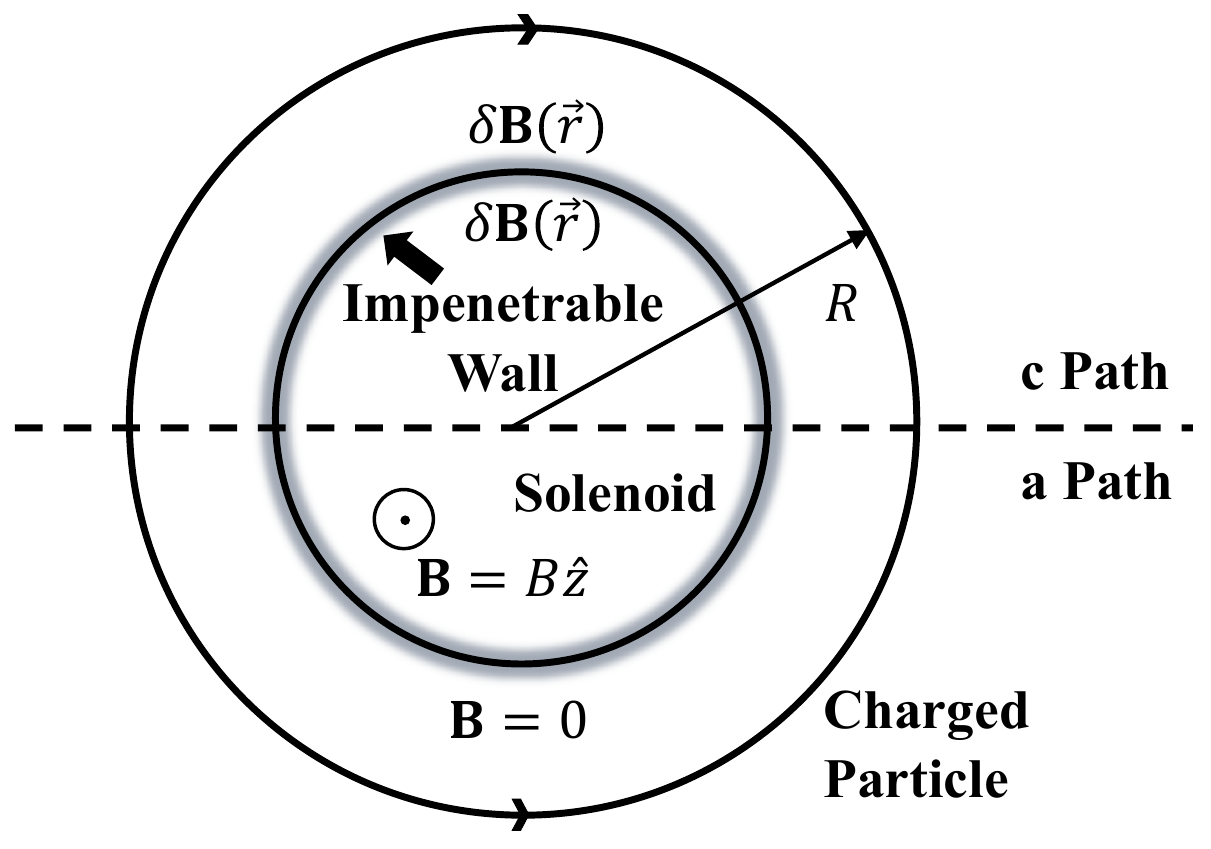}
			\caption{ A thin cylindrical solenoid of infinite length is very closely wound so that the magnetic field is essentially confined within the solenoid. The charged quantum particle is moving in a circle of radius $R$. }	\label{fig2}
		\end{center}
	\end{figure}

	As a simplified model shown in Fig.~\ref{fig2}, we imagine that the charged particle is moving in a circle of radius $R$.
	When the particle is at position $\mathbf{r}^\prime$, the field it generates at position $\mathbf{r}$ follows the Biot-Savart law:
	\begin{eqnarray}
	\label{deltaB}
	\delta\mathbf{B}=\frac{\mu_0}{4\pi }\cdot\frac{q\mathbf{v}\times(\mathbf{r-r^\prime})}{\abs{\mathbf{r-r^\prime}}^3}
	\end{eqnarray}
	We further assume that the solenoid is very thin, so that we need to consider only $\delta\mathbf{B}$ at the center of the solenoid.
	\begin{eqnarray}
	\label{deltaBatSolenoid}
	\begin{aligned}
	\delta B_z(r=0,z)=\frac{\mu_0qR^2}{2T(R^2+z^2)^{\frac{3}{2}}}
	\end{aligned}
	\end{eqnarray}
	So the phase difference is 
	\begin{eqnarray}
	\label{case1}
	\begin{aligned}
	\Delta\phi&=\frac{\Delta S}{\hbar}\\
	&=\frac{1}{\mu_0\hbar}\int_0^Tdt\iiint dxdydz\;\mathbf{B}\cdot\delta\mathbf{B}\\
	&=\frac{1}{\mu_0\hbar}\int_0^Tdt\iiint dxdydz\;B_z\delta B_z\\
	&=\frac{A}{\mu_0\hbar}\int_0^Tdt\int dz\;B_z\delta B_z\\
	&=\frac{\Phi_{\mathbf{B}}}{\mu_0\hbar}\int_0^Tdt\int dz\;\delta B_z\\
	&=\frac{q\Phi_{\mathbf{B}}}{2\hbar}\int_{-\infty}^{+\infty} dz\;\frac{R^2}{(R^2+z^2)^{\frac{3}{2}}}\\
	&=\frac{q\Phi_{\mathbf{B}}}{\hbar}
	\end{aligned}
	\end{eqnarray}
	So we got  the Aharonov-Bohm phase, without using a vector potential,  by expressing the Lagrangian in terms of the gauge invariant magnetic field in all space.
	
	\subsection{A finite closed toroidal coil}
\begin{figure}
		\setlength{\belowcaptionskip}{-0.5cm}
		\vspace{-0.1cm}
		\begin{center}
			\includegraphics[width=5cm]{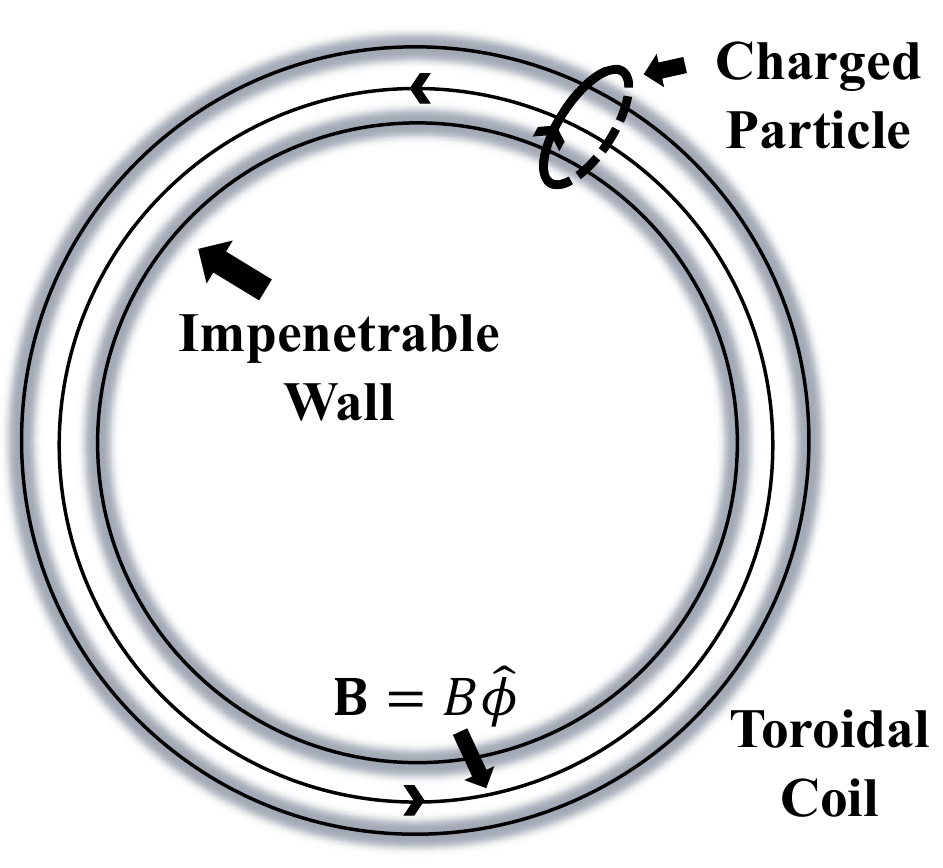}
			\caption{ A thin toroidal  coil  consists  of  a  circular  ring around  which  a  long  wire  is  wrapped. The winding is uniform and tight. Phase difference is calculated between two paths, which encircle the coil, of the charged particle.}
			\label{fig3}
		\end{center}
	\end{figure}

	A solenoid of infinite length can never be realized in an experiment, so in this section we  calculate the Aharonov-Bohm phase for the realistic case of a finite closed magnetic flux tube. 
	As shown in Fig.~\ref{fig3}, we take a toroidal coil consisting of a circular ring, or ``donut", around which a long wire is wrapped uniformly and tight enough so that each turn can be considered a closed loop. The magnetic field is circumferential at all points inside the coil and zero outside. Again assuming that the coil is very thin, so that we can ignore the axial distance dependence of the field, we have
	\begin{eqnarray}
	\mathbf{B}=\left\{ 
	\begin{array}{ll}
	B\hat{\mathbf{\theta}}\quad &\text{inside the coil}\\
	0\quad  & \text{outside the coil}
	\end{array}
	\right.
	\end{eqnarray}
	Now the phase difference is 
	\begin{eqnarray}
	\label{case2}
	\begin{aligned}
	\Delta\phi&=\frac{\Delta S}{\hbar}\\
	&=\frac{1}{\mu_0\hbar}\int_0^Tdt\iiint d\rho d\theta dz\;\rho\mathbf{B}\cdot\delta\mathbf{B}\\
	&=\frac{A}{\mu_0\hbar}\int_0^Tdt\int_{0}^{2\pi} d\theta\;\rho B\delta B_{\theta}\\
	&=\frac{\Phi_{\mathbf{B}}}{\mu_0\hbar}\int_0^Tdt\int_{0}^{2\pi} d\theta\;\rho \,\delta B_{\theta}\\
	&=\frac{\Phi_{\mathbf{B}}}{\mu_0\hbar}\int_0^Tdt\oint d\mathbf{l}\cdot \delta\mathbf{B}\\
	&=\frac{\Phi_{\mathbf{B}}}{\mu_0\hbar}\int_0^Tdt\;\mu_0 I(t)\quad\text{(Amp\`{e}re's law)}\\
	&=\frac{q\Phi_{\mathbf{B}}}{\hbar} \, ,
	\end{aligned}
	\end{eqnarray}
	which is the same result as in the last section.
	
	\subsection{ A solenoid with open ends}
\begin{figure}
		\setlength{\belowcaptionskip}{-0.1cm}
		\vspace{-0.1cm}
		\begin{center}
			\includegraphics[width=6cm]{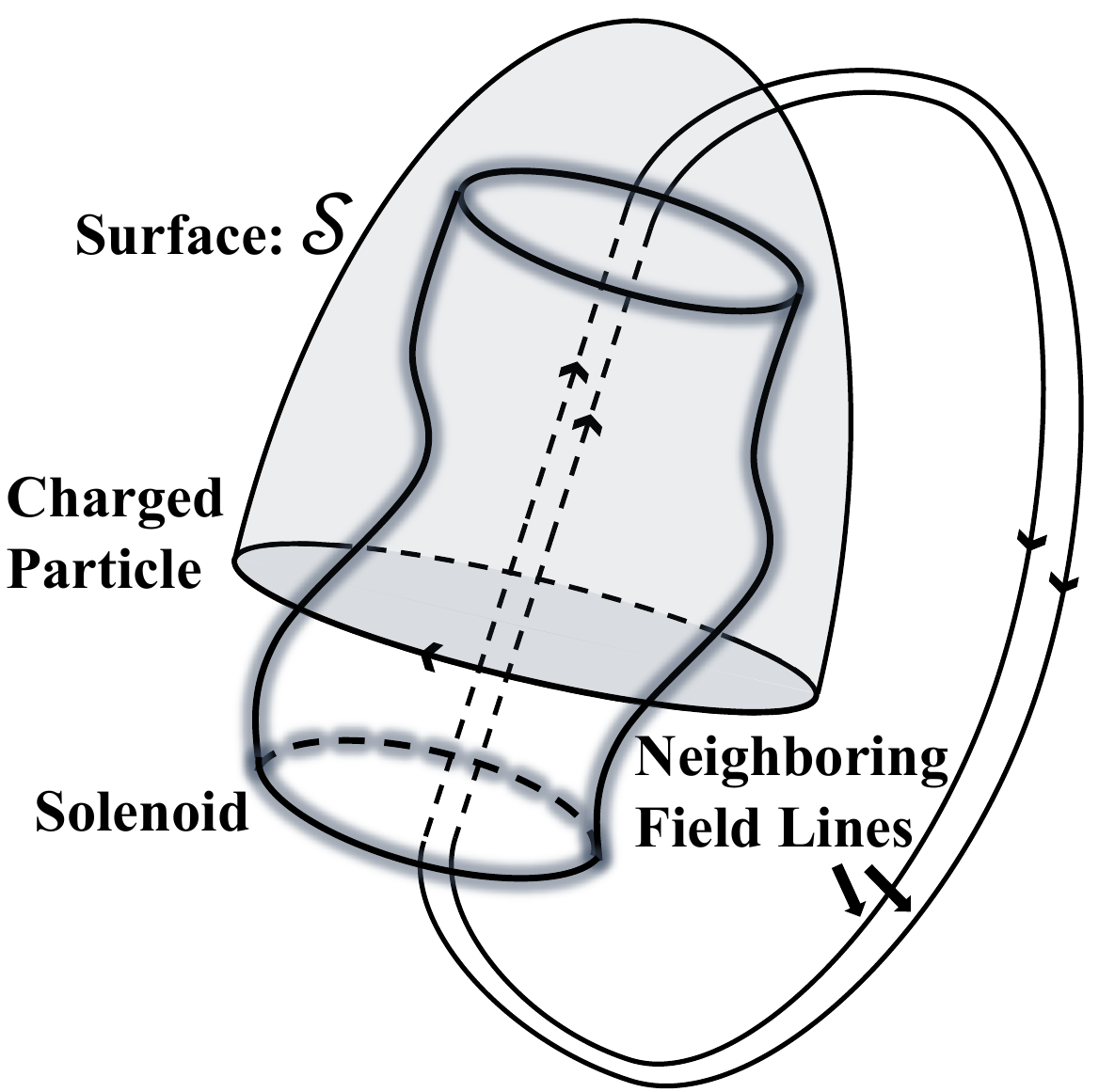}
			\caption{A source, a solenoid for example, generates magnetic field. By dividing the space into shells using neighboring surfaces defined by neighboring magnetic field lines and by further dividing the shells into thin rings, we calculated the phase difference between two paths, which form a closed loop, of the charged particle. This is the most general calculation for the Aharonov-Bohm phase.}
			\label{fig4}
		\end{center}
	\end{figure}

	In the last two sections, the magnetic field was completely inaccessible to the charged particle, so the overall space was not simply connected,  and we also considered very simple symmetric geometries. 
	In this section we relax these conditions and consider a general situation where the magnetic field extends into the region where the charged particle moves, and the solenoid can have a more complicated shape, as indicated in Fig.~\ref{fig4}. 
	Moreover, the charged particle can move in an arbitrary trajectory $\mathbf{x}(t)$, as long as it is a loop enclosing non-zero magnetic flux.
	To do the integration, we divide the space into shells between two neighboring surfaces defined by neighboring magnetic field lines (shown in  Fig.~\ref{fig4}) and further divide these shells into many small rings by cutting the field lines perpendicularly. The relevant AB phase becomes,  
	\begin{eqnarray}
	\label{case3}
	\begin{aligned}
	\Delta\phi&=\frac{\Delta S}{\hbar}\\
	&=\frac{1}{\mu_0\hbar}\int_0^Tdt\iiint dxdydz \;\mathbf{B}\cdot\delta\mathbf{B}\\
	&=\frac{1}{\mu_0\hbar}\int_0^Tdt\sum_{shells}\sum_{rings} \Delta A\Delta l\,\mathbf{B}\cdot\delta\mathbf{B}\\
	&=\frac{1}{\mu_0\hbar}\int_0^Tdt\sum_{shells}\sum_{rings} \Delta A\Delta l\,B\delta B_l\\
	&=\frac{1}{\mu_0\hbar}\int_0^Tdt\sum_{shells}\Delta\Phi\sum_{rings} \Delta l\,\delta B_l\\
	&=\frac{1}{\mu_0\hbar}\sum_{shells}\Delta\Phi_{\mathbf{B}}\int_0^Tdt\sum_{rings} \Delta l\,\delta B_l\\
	&=\frac{1}{\mu_0\hbar}\sum_{shells}\Delta\Phi_{\mathbf{B}}\int_0^Tdt\;\mu_0 I(t)\quad\text{(Amp\`{e}re's law)}\\
	&=\frac{q}{\mu_0\hbar}\sum_{shells}\Delta\Phi_{\mathbf{B}}^{enc}
	=\frac{q\Phi_{\mathbf{B}}^{enc}}{\hbar}
	\end{aligned}
	\end{eqnarray}
	
	Here $\Delta A$ is the area of a ring, $\Delta l$ is a line segment along the flux line, and $B_l$ is the component of the magnetic field along the flux line. 
	Note that only those closed flux lines that encircles the path of the charged particle contributes to $\int I(t)dt$. Thus, in this situation, where the solenoid is long compared with the size of the charged particle's loop, very few flux lines will not intersect the loop and the AB phase will be essentially the same as for an infinite solenoid. Also note that the result in Eq.~\eqref{case3} for the AB phase is simply obtained in the local formulation by using the Stokes theorem on the surface $\mathcal S$ shown in  Fig.~\ref{fig4} to calculate $W[C]$. 
	
	
	\section{V. The Electric AB Effect: Vaidman's One-Dimensional Gedanken  Interferometer}
	
	Although not as well known as the original magnetic AB effect, there is also a corresponding electric effect, as would be expected in a relativistic context.
	We will analyze this electric AB effect from our perspective using the one-dimensional ``gedanken'' interferometer experiment proposed by Vaidman~\cite{PhysRevA.86.040101}. 
	
\begin{figure}
		\setlength{\belowcaptionskip}{-0.5cm}
		\vspace{-0.1cm}
		\begin{center}
			\includegraphics[width=6.5cm]{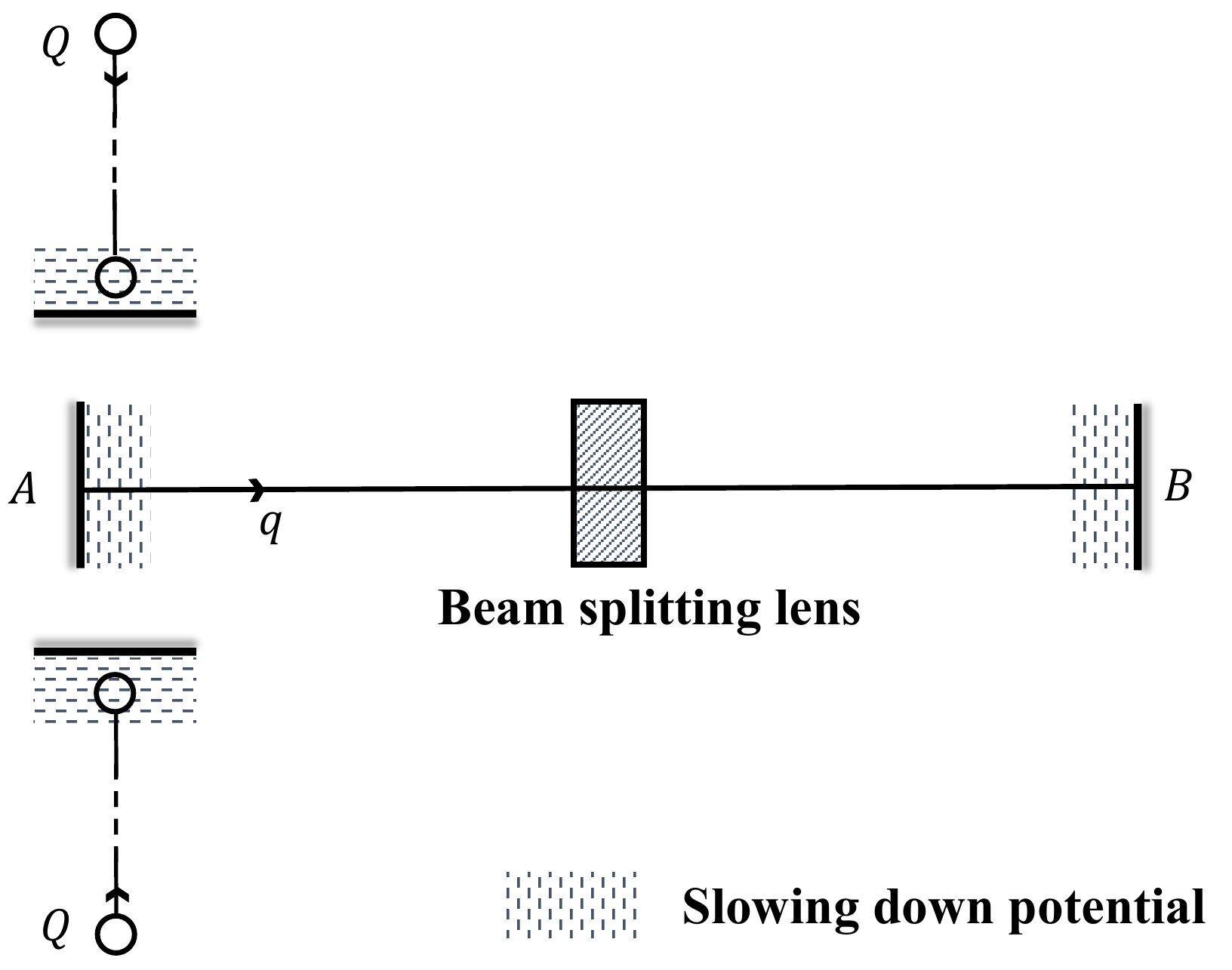}
			\caption{ Vaidman's one-dimensional gedanken interferometer. }	\label{fig5}
		\end{center}
	\end{figure}

	Figure~\ref{fig5} illustrates the interferometer in which a particle with charge $q$ travels between two mirrors A and B with a beam-splitting lens between them.
	The system is first tuned so that the wave function of the particle interferes destructively at mirror A, but  reaches mirror B with certainty.
	Next, a potential is designed around each mirror to slow down the particle, so that it spends a long time $\tau$ near the mirrors. 
	The range of this potential is assumed to be very small compared to the size of the whole interferometer.
	
	The electric AB phase is now generated by two extra charges $Q$ placed symmetrically on the perpendicular axis at equal distances from mirror A. 
	When the charge $q$ is far away from A, the charges $Q$ are very far from the whole interferometer so that their potential can be neglected. 
	However, when the charge $q$ moves into the ``slowing down potential'' region near A, the charges $Q$ are quickly moved close to mirror A, and are then 
	slowed down by another potentials so that they spend a rather long time $T$ with $T < \tau$ near A. 
	
	The charges $Q$ do not result in any electrostatic force on the particle $q$, since, because of the symmetric arrangement, the electric field $\mathbf{E}$ always vanishes at the position of the particle. This is not true for the electrostatic \emph{potential} which  during the time $T$, is given by $A_0^{(Q)}(\mathbf{r}_q)=2\times\frac{1}{4\pi\epsilon_0}\frac{Q}{r}$, where $r$ denotes the distance between charge $Q$ and charge $q$.
	Thus, because of the term $-q A_0$ in Eq.~\eqref{Le+LeF}, there is an electric AB phase~\cite{PhysRevA.86.040101}
	\begin{eqnarray}
	\label{electricAB1}
	\begin{aligned}
	\Delta \phi =- \frac{qA_0^{(Q)}(\mathbf{r}_q)T}{\hbar}= -\frac{1}{4\pi\epsilon_0}\frac{2qQT}{r\hbar}.
	\end{aligned}
	\end{eqnarray}
	
	This phase shift affects the interference of the direct and reflected waves in the interferometer, so there is now a finite probability to for the particle to be close to  A. 
	Thus, the AB effect can be observed by detecting a charge $q$ at mirror A. 
	
	The point is now that by using Eq.~\eqref{model1} or Eqs.~\eqref{heuren1}-Eq.~\eqref{heuren4} we can exactly rewrite the  AB phase Eq.~\eqref{electricAB1} as
	\begin{eqnarray}
	\label{electricAB2}
	\begin{aligned}
	\Delta \phi =  - \epsilon_0\int d\mathbf {x}\,  \mathbf{E}_Q \cdot \delta\mathbf{E} (\mathbf {x}),
	\end{aligned}
	\end{eqnarray}
	where $\mathbf{E}_Q$ is the electric field due to the charges $Q$ and $\delta\mathbf{E}$ is the field due to the charge $q$.
	Thus, we have obtained the correct electric AB phase using only gauge independent electric fields.
	
	
	\section{VI. Discussion, summary and outlook}
	
	The  derivation of AB phases given in this paper sheds light on the questions related to gauge invariance, and locality versus non-locality in quantum mechanics.
	In particular, we demonstrated that in quasi-stationary situations, AB phases can be recovered using only gauge invariant fields, but at the expense of giving up locality. Our discussion was in terms of path integrals.
	Note that by including the Maxwell term Eq.~\eqref{LF} in the path integral, we indeed treated the electromagnetic field as part of the \emph{quantum} system, although we considered only a quasi-static process where radiation can be neglected.
	We also remark that as long as radiation can be neglected, inclusion of time variation would not alter our conclusion. 
	Moreover, since our analysis involved only the energies, it could equally well have been  made in  an Hamiltonian framework where the AB phase shift emerges as a Berry phase\cite{berry}.  
	
	We now discuss some relevant previous works and compare them to ours.
	
	1) \emph{Local force on the source versus non-local field.} \\
	As shown in Sections. IV and V, our description gives results consistent  with Vaidman's~\cite{PhysRevA.86.040101}.
	His gauge-independent description was based on \emph{local forces} on the sources, and so was Kang's ~\cite{PhysRevA.91.052116}.
	These results were challenged by Aharonov and collaborators using examples in which an AB phase is present without any local force on the sources ~\cite{PhysRevA.92.026101}.
	Our claim instead focuses on the non-local effect of the field energy, which is related to the source but does not necessarily imply local force on the sources.
	An AB phase without any local fields at the position of the test particle, and an AB phase without any local field at the source, are in essence two sides of the same coin.
	This is very natural since the electromagnetic energy of the source and the test particle correspond one to one, as made clear from Eq.~\eqref{heuren1}-Eq.~\eqref{heuren5}.
	
	2)  \emph{Necessity of gauge dependent description}\\
	Our analysis suggests that in general gauge potentials are still needed in order to give a local description.
	Indeed,  descriptions based on quantizing the electromagnetic field, such as Refs.~\cite{PhysRevLett.125.040401, PhysRevA.104.032219, Saldanha}, do need the local gauge dependent potentials.
	
	3) \emph{Validity of a field-based description.}\\
	The field-based description was also questioned in Ref.~\cite{PhysRevA.92.026101}, but in Section. III, we provided a formal argument in support of our picture.
	
	4)  \emph{Can the AB phase be attributed to the electromagnetic field?}\\
	Pearle and Rizzi~\cite{PhysRevA.95.052123, PhysRevA.95.052124} showed that the AB phase can be attributed to one of three entities: the electron (test particle), the solenoid (source), or the vector potential (electromagnetic field).
	In our description the AB phase can indeed be attributed to the electromagnetic field, and it is thus consistent with their findings.
	This conclusion is also consistent with Refs.~\cite{PhysRevA.104.032219, Saldanha}, where the AB phase was attributed to the change in the electromagnetic vacuum energy due to the local interaction between photons and the test particle. 
	
	We also note that  Marletto and Vedral~\cite{PhysRevLett.125.040401} proposed that the AB phase is gradually acquired along the test particle's path.
	Our path integral description is, at least conceptually,  consistent with their finding, but to test it one would have  to measure the phase difference at different points along the particle's trajectory. 
	
	To summarize, in the works where local descriptions of the AB effect have been proposed, locality significantly relies on either the use of classical (external) vector potential, like in Refs.~\cite{PhysRev.115.485, PhysRevD.12.3845}, or the use of a quantized (dynamical) vector potential, like in Refs.~\cite{PhysRevA.95.052124, PhysRevLett.125.040401, PhysRevA.104.032219, Saldanha}. 
	On the other hand, \emph{local} gauge-invariant approaches like in Refs.~\cite{PhysRevA.86.040101,PhysRevA.91.052116} were criticized  in Refs.~\cite{PhysRevA.92.026101, PhysRevA.93.042110}, and Ref.~\cite{PhysRevLett.125.040401} also pointed out that there is a residual non-locality hidden in the non-local correspondence between the phase of the source and the phase of the test particle.
	Our formulation makes it clear that a gauge independent description is still available, but at the expense of sacrificing locality.
	
	Our non-local gauge independent formulation might be of help to describe  AB effects in a relativistic context, which must employ the full theory of  quantum electrodynamics.
	Since it employs only gauge invariant quantities, it might in particular provide a natural way to understand how the AB effect is compatible with causality.
	This issue, while beyond the scope of present paper, is both important and fundamental and most certainly deserves further investigation.
	
	\section{Acknowledgments}
	We thank A. Hegg for useful discussion. 
	This work is supported by National Natural Science Foundation of China No. 12274287 and No. 12042507 and Innovation Program for Quantum Science and Technology No. 2021ZD0301900.
	
	\section{Appendix}
	In this appendix, we show that using sources as variables, the appropriate Lagrangian for the field to use is Eq.~\eqref{LF}. 
	
	Usually, the Lagrangian density of the electromagnetic field is written as a function of the electromagnetic potentials, while the sources are considered fixed. We are interested in the variation of the Lagrangian,
	\begin{eqnarray}
	\begin{aligned}
	\delta \mathcal L\left(A_\mu,\partial_{\nu}A_\mu\right)=\delta \left(\frac{\epsilon_0}{2}\mathbf{E}^2-\frac{1}{2\mu_0}\mathbf{B}^2\right)-\sum_{\mu=0}^{3}J^\mu dA_\mu\\
	=\sum_{\mu,\nu=0}^{3}\frac{\partial\left(\frac{\epsilon_0}{2}\mathbf{E}^2-\frac{1}{2\mu_0}\mathbf{B}^2\right)}{\partial(\partial_{\nu}A_\mu)}\delta (\partial_{\nu}A_\mu)-\sum_{\mu=0}^{3}J^\mu dA_\mu
	\end{aligned}
	\end{eqnarray}
	where we defined the four-position, four-potential, and four-current-density as
	\begin{align}
		x^\mu&=\left(ct,x,y,z\right)\\
		A^\mu&=\left(\frac{A_0}{c},\mathbf{A}\right)\\
		J^\mu&=\left(c\rho,\mathbf{J} \right)
	\end{align}
	where $c=\frac{1}{\sqrt{\epsilon_0\mu_0}}$ is the speed of light.
	The electric field $\mathbf{E}$ and the magnetic field $\mathbf{B}$ are defined via the first derivatives of the potentials
	\begin{align}
		\mathbf{E}&=-\gradient A_0-\frac{\partial\mathbf{A}}{\partial t}\\
		\mathbf{B}&=\curl\mathbf{A}
	\end{align}
	Define the action as 
	\begin{align}
		\mathcal{S}=\int d^4x\,\mathcal{L}\left(A_\mu,\partial_{\nu}A_\mu\right)
	\end{align}
	The least action principle leads to the Euler-Lagrangian equation:
	\begin{align}
		\frac{\partial\mathcal{L}}{\partial A_\mu}-\sum_{\nu=0}^3\frac{d}{dx_\nu}\left(\frac{\partial\mathcal{L}}{\partial\left(\partial_{\nu}A_\mu\right)}\right)=0
	\end{align}
	which leads to Maxwell's equations
	\begin{align}
		\divergence\mathbf{E}&=\frac{\rho}{\epsilon_0}\\
		\curl\mathbf{B}&=\mu_0\left(\mathbf{j}+\frac{\partial\mathbf{E}}{\partial t}\right)
	\end{align}
	
	However, in the Lagrangian of the charged particle Eq.~\eqref{Le+LeF}, we are keeping $\mathbf{x}$ and $\dot{\mathbf{x}}$, i.e. $J_\mu$ and $\partial_\nu J_\mu$, as variables, instead of $A_\mu$ and $\partial_{\nu}A_\mu$.
	To get the appropriate Lagrangian of the field, we need to perform Legendre transformation to change variables.
	Define
	\begin{align}
		\Tilde{\mathcal{L}}\left(J_\mu,\partial_\nu J_\mu\right)=-\sum_{\mu=0}^{3}J^\mu A_\mu-\mathcal L\left(A_\mu,\partial_{\nu}A_\mu\right)
	\end{align}
	We have
	\begin{eqnarray}
	\begin{aligned}
	\delta \Tilde{\mathcal{L}}\left(J_\mu,\partial_\nu J_\mu\right)&=-\sum_{\mu=0}^{3}J^\mu \delta A_\mu-\sum_{\mu=0}^{3}A^\mu \delta J_\mu-d\mathcal{L}\\
	&=\delta \left(\frac{1}{2\mu_0}\mathbf{B}^2-\frac{\epsilon_0}{2}\mathbf{E}^2\right)-\sum_{\mu=0}^{3}A^\mu dJ_\mu
	\end{aligned}
	\end{eqnarray}
	Note that the minus sign in front of $\sum A_\mu dJ_\mu$ matches the sign of $L_{int}$, so $\Tilde{\mathcal{L}}$ is the proper Lagrangian density to put in the path integral formulation, which differs from the usual Lagrangian density $-\frac{1}{4\mu_0} F_{\mu\nu}F^{\mu\nu}$ by a sign.
	
	\bibliography{Gauge_Independent_Description_of_AB_Effect}
\end{document}